\documentclass[%
 reprint,
 amsmath,amssymb,
 aps,
pre,
notitlepage
]{revtex4-1}
\usepackage[ruled,vlined]{algorithm2e}
\usepackage{graphicx}% Include figure files
\usepackage{dcolumn}% Align table columns on decimal point
\usepackage{bm}% bold math
\usepackage{float}
\begin{document}

\title{Fast and simple decycling and dismantling of networks}

\author{Lenka Zdeborov\'a}
\affiliation{
Institut de Physique Th\'erique, CNRS, CEA and Universit\'e Paris-Saclay, Gif-sur-Yvette, France
}
\author{Pan Zhang and Hai-Jun Zhou}
\affiliation{
CAS Key Laboratory of Theoretical Physics, Institute of Theoretical Physics, Chinese Academy of Sciences, Beijing 100190, China
}

\date{\today}

\begin{abstract}
Decycling and dismantling of complex networks are underlying many
important applications in network science. Recently these two closely
related problems were tackled by several heuristic algorithms, simple and considerably sub-optimal, on the one hand, and time-consuming message-passing ones that 
evaluate single-node marginal probabilities, on the other hand. In this paper we
propose a simple and extremely fast algorithm, CoreHD, which
recursively removes nodes of the highest degree from the $2$-core of the network. CoreHD performs much better than all existing simple
algorithms. When applied on real-world networks, it
achieves equally good solutions as those obtained by the state-of-art 
iterative message-passing algorithms at greatly reduced
computational cost, suggesting that CoreHD should be the algorithm of choice 
for many practical purposes. 
%This algorithm can be easily generalized to decycling and dismantling defined on $k$-core of networks.
\end{abstract}

\maketitle

%\tableofcontents

\section{\label{sec:level1}Introduction}

In decycling of a network we aim to remove as few nodes as possible 
such that after the removal the remaining network contains no loop.
In network dismantling we aim to find the smallest set of nodes such
that after their removal the network is broken into connected 
components of sub-extensive size. These are two fundamental 
network-optimization problems with a wide range of 
applications, related to optimal vaccination and 
surveillance, information spreading, viral marketing, and identification of
influential nodes. 
Considerable research efforts have been devoted to the network 
decycling and dismantling problems recently~\cite{zhou2013spin,altarelli2013optimizing,guggiola2015minimal,morone2015influence,mugisha2016identifying,braunstein2016network,qin2016spin,Clusella-etal-2016}.

Both the decycling and the dismantling problems belong to the 
class of NP-hard problems~\cite{karp1972reducibility,braunstein2016network}, meaning
that it is rather hopeless to look for algorithms to solve them 
exactly in polynomial time. However,  finding the best
possible approximate solutions for as large classes of networks as 
possible is an open and actively investigated direction. Recent theoretic and algorithmic progress on both these problems~\cite{zhou2013spin,altarelli2013optimizing,guggiola2015minimal,mugisha2016identifying,braunstein2016network}
came from the fact that, on random sparse networks with degree 
distributions having a finite second moment, methods from physics of spin glasses provide accurate algorithms for both decycling and dismantling. 
These sparse random networks are locally tree-like and do not 
contain many short loops. On such networks the decycling is closely 
linked to dismantling and asymptotically almost the same set of nodes 
is needed to achieve 
both~\cite{janson2008dismantling,mugisha2016identifying,braunstein2016network}.
Even on real-world networks that typically contain many small loops, 
best dismantling is currently achieved by first finding a decycling
solution and then re-inserting nodes that close short loops but do
not increase too much the size of the largest 
component~\cite{mugisha2016identifying,braunstein2016network}.

Both the algorithms of \cite{braunstein2016network} and \cite{mugisha2016identifying} achieve performance that is extremely 
close to the theoretically optimal values computed on random
networks. However, both these algorithms are global, they need to 
iterate certain equations on the whole network in order to select the
suitable candidate nodes. Although they are both scalable and can be 
run on networks with many millions of nodes, they are not completely 
straightforward to understand and require some experience with spin glass theory. The close-to-optimal performance of these
algorithms is theoretically justified only on 
random networks. Despite their good performance observed empirically
on networks with many loops, there  might still exist even better and analyzable strategies for real-world networks.

As usual in applied science, in many potential applications we are at 
first not even sure that optimal dismantling or optimal decycling is 
the best strategy to answer the question in hand (e.g., the problem of social influence maximization~\cite{Richardson-Domingos-2002,Kempe-etal-2015,jung2012irie,borgs2014maximizing}). 
Therefore it is extremely important to have a really very simple and 
fast decycling and dismantling strategy that can provide an accurate assessment of whether this approach is at all interesting 
for a given practical problem.
However, existing simple strategies, such as removing adaptively high 
degree nodes~\cite{albert2000error,cohen2001breakdown},
are very far from optimal performance and therefore not very useful.
Recently the authors of \cite{morone2015influence} claimed that a heuristics based on the so-called {\it collective influence} (CI) measure can be a perfect candidate for this purpose. This algorithm has attracted a lot of enthusiasm in the network science community.  However, more systematic investigations performed in  \cite{mugisha2016identifying,braunstein2016network,Clusella-etal-2016} revealed that the CI algorithm is still annoyingly far from being optimal. The CI algorithm is also not particularly competitive in terms of computational time because a large neighborhood of a node needs to be considered in order to evaluate the CI measure. 

In the present paper we introduce the CoreHD algorithm that is basically as simple and fast as the adaptive removal of high degree nodes, yet its performance is much closer to optimal than the CI algorithm or its extended versions,
and comparably close as the best known message-passing methods \cite{mugisha2016identifying,braunstein2016network} while several orders of magnitude faster. It hence provides simple and tractable solutions for networks with many billions of nodes. The method is simply based on adaptive removal of highest-degree nodes from the $2$-core of the network. 
Apart of its simplicity and speed the performance of the CoreHD algorithm is basically indistinguishable from the performance of the message-passing algorithms on random 
graphs with long-tailed degree distributions. 
On all real-world network instances we tested the result by CoreHD is
within few nodes from the best one found by message-passing and on some instances we found that it is even slightly better. 
On top of all that, the simple structure of CoreHD might be amenable to rigorous analysis providing guarantees for loopy networks that are not accessible for the message-passing methods. 

For all the above reasons we argue that in many applications of decycling and dismantling CoreHD should be the first choice.
The simple algorithmic idea  generalizes easily to the problem of destroying optimally the $k$-core of a network - one focuses on the current $k$-core and adaptively removes highest degree nodes.

\section{The CoreHD Algorithm}

We now describe CoreHD as an extremely fast algorithm for decycling and dismantling of huge complex networks with close-to-optimal 
outcomes. Let us begin with some motivating discussions.

\begin{figure}[h]
   \centering
   \includegraphics[width=0.7\columnwidth]{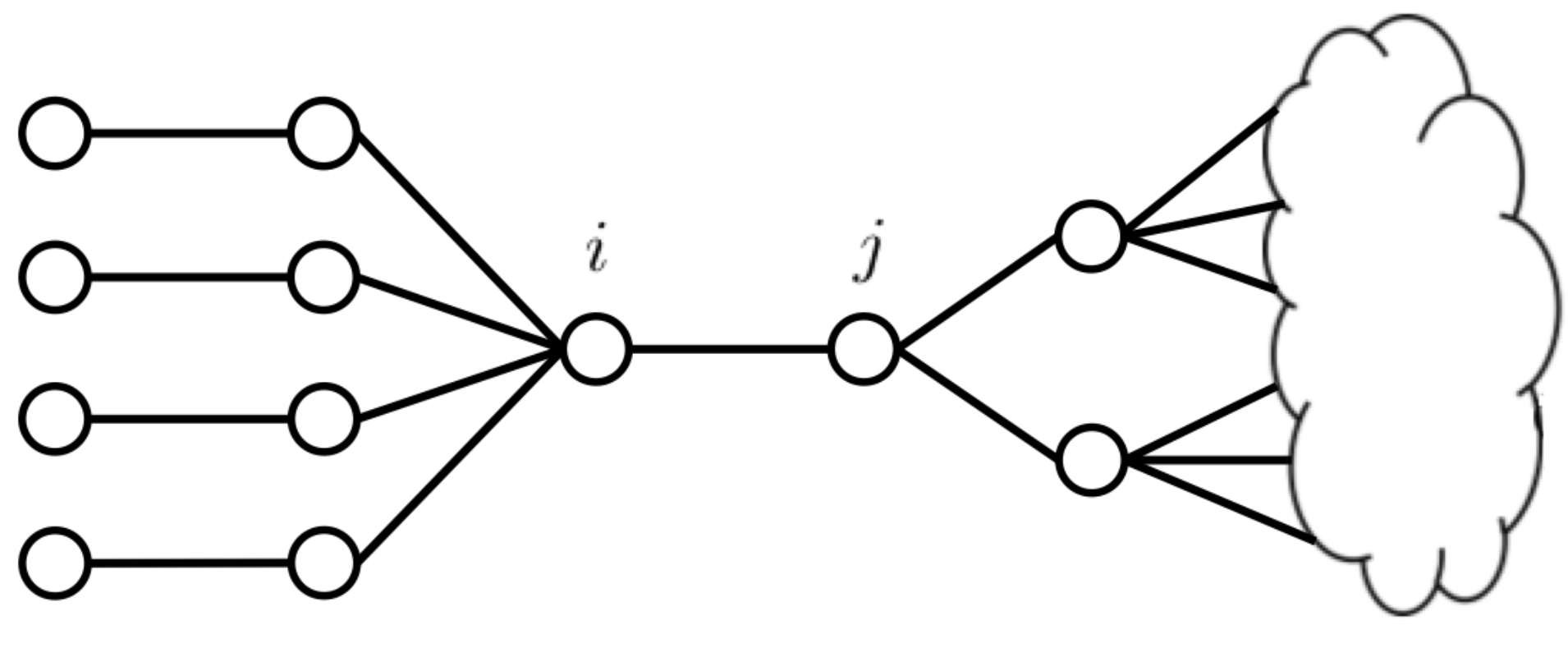}
   \caption{Illustration of a network with dangling trees. Each circle denotes a node in the network, each line connecting circles denotes an edge, and the cloud represents the other part (nodes and edges) of the network. \label{fig:dtree}}
\end{figure}

Perhaps the simplest algorithms one can propose for decycling and dismantling is adaptive removal of highest-degree nodes.  We call this method HD, it is indeed extremely fast, but empirically does not perform very well. One reason why HD 
does not work well is that some nodes of large degree, such as node $i$ in Fig.~\ref{fig:dtree}, do not belong to any loop, and hence do not have to be removed for decycling. Due to the property that trees can always be dismantled by a vanishing fraction of nodes~\cite{janson2008dismantling}, nodes such as $i$ of Fig.~\ref{fig:dtree} 
are also not useful for dismantling. Note that the CI method of \cite{morone2015influence} shares this problem, see the appendix. 

The above observation motivates a very  natural idea that dismantling and decycling algorithms should always focus only on the $2$-core of 
the network. The $2$-core is a sub-network that is obtained after adaptive removal of all leaves (nodes with only a single attached edge). 
The simplest and fastest strategy is then to remove the highest-degree nodes from the remaining $2$-core.
To our surprise this simple idea provides much better performance than other comparably simple approaches existing in the literature. We call the resulting algorithm CoreHD, it is
detailed in Algorithm~\ref{CoreHD}.

\begin{algorithm}
\caption{CoreHD} 
\label{CoreHD}
	\KwIn{A network.}
	 \KwOut{A forest of small trees.}
\begin{enumerate}
	\item Find the $2$-core of the network, and obtain the degree of every node within this $2$-core (edges to outside nodes not considered).
    \item Identify the node $i$ with the largest degree in the 
    $2$-core. If there are more nodes with the same largest degree, randomly choose one of them.
	\item Remove node $i$, update the $2$-core and the degrees of all its nodes. If the $2$-core is empty, then do tree-breaking and 
    stop; otherwise go to step $2$.
\end{enumerate}
\end{algorithm}

For the decycling problem, CoreHD simply removes highest-degrees 
nodes from the $2$-core in an adaptive way (updating node degree
as the $2$-core shrinks), until the remaining network becomes a forest. For dismantling, after decycling, CoreHD also
breaks the trees into small components, see Appendix that follows tree-breaking strategy from \cite{mugisha2016identifying,braunstein2016network}. 
In case the original network has many small loops, a refined
dismantling set is obtained after a reinsertion of nodes that
do not increase (much) the size of the largest component, 
again as proposed recently in~\cite{mugisha2016identifying,braunstein2016network}.
For details on implementation of the  reinsertion algorithm we refer to the Appendix.

\section{Results}

In this section we evaluate the CoreHD algorithm for both random and real-world networks, by comparing the minimum fraction of nodes we need to remove in order to break the network
into a forest or components with size smaller than $0.01n$.
We compare to the Belief Propagation guided Decimation (BPD) \cite{mugisha2016identifying} and Collective Influence method (CI) \cite{morone2015influence} (CI$_4$ results are obtained using 
the original code of Ref.~\cite{morone2015influence}).

First, we notice that on some simple examples, e.g. regular random graphs with degree 
$3$, the CoreHD algorithm reaches the exact optimal decycling fraction $\rho=0.25$. This matches the performance of a greedy method of \cite{bau2002decycling} that for this particular case is provably optimal.  %This is because CoreHD always removes nodes with degree $3$ when they exist. After arriving at a critical network where every node has degree $2$ (by removing $0.25 N$ nodes), CoreHD then breaks the loops and chains sharply.

\begin{figure}[t]
   \centering
   \includegraphics[width=0.7\columnwidth]{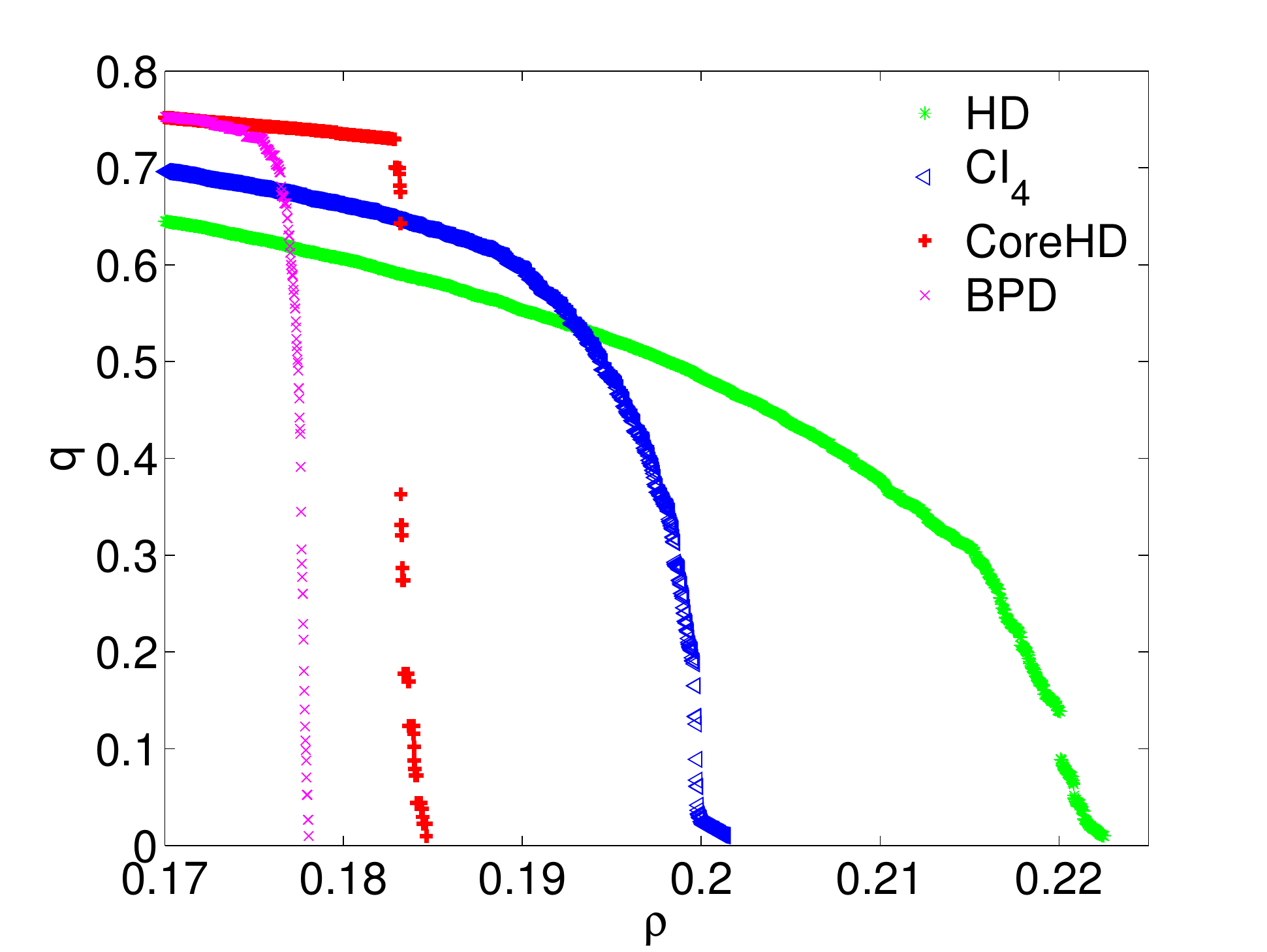}
   \includegraphics[width=0.7\columnwidth]{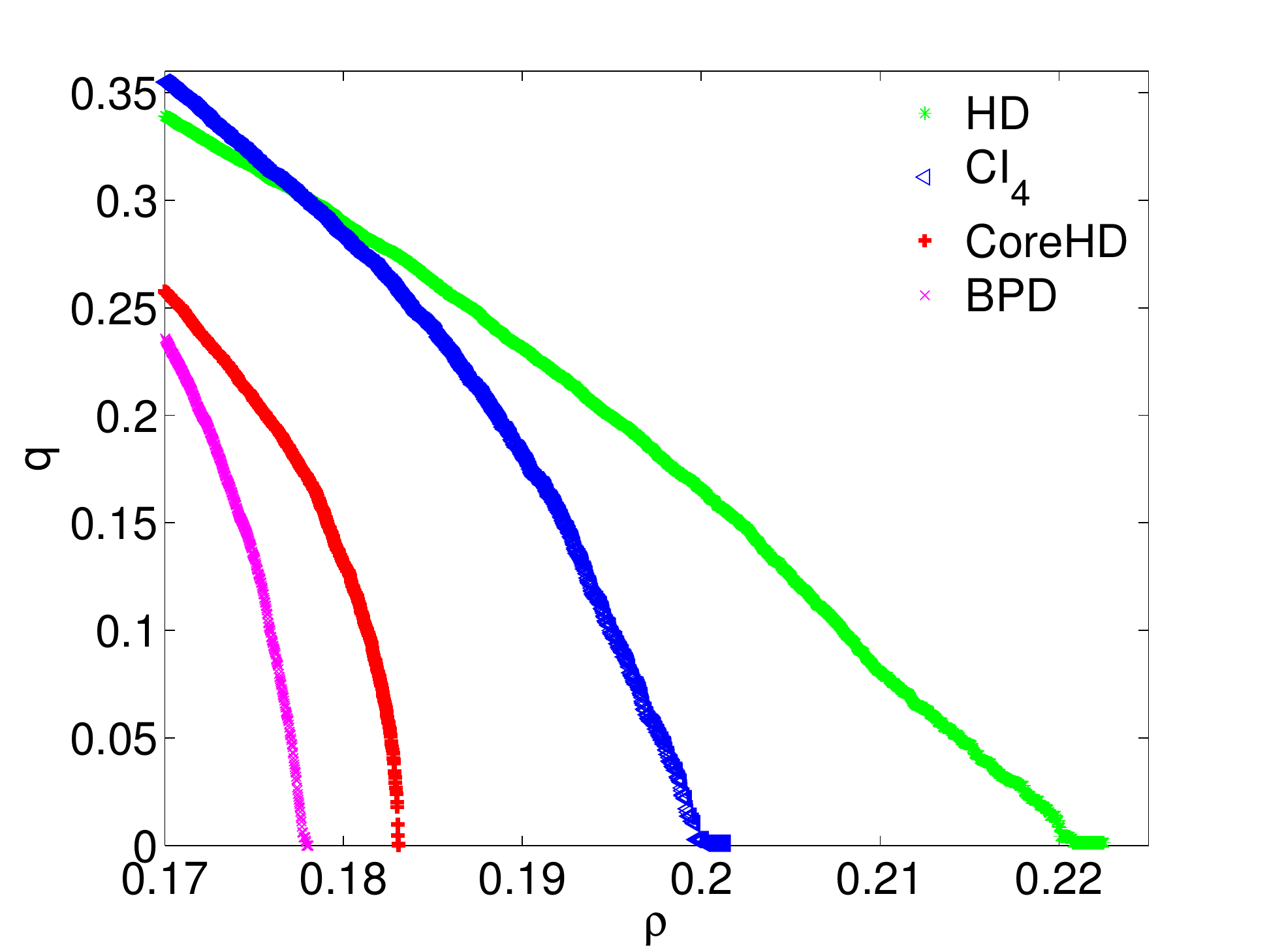}
   \caption{Fraction of nodes in the largest connected component (LCC) \textit{(upper)} and in  the $2$-core \textit{(lower)} as a function
   of fraction of nodes removed, for HD, {CI}$_4$, CoreHD and BPD on an Erd\H{o}s-R\'enyi random graph with 
   number of nodes $N=5\times 10^4$, and average degree $c=3.5$. In all four methods 
   nodes are removed one by one. \label{fig:ovl}}
\end{figure}

In Fig.~\ref{fig:ovl} we compare the performance of the above algorithms on an Erd\H{o}s-R\'enyi random network with $N=50000$ nodes and 
average degree $c=3.5$. In the upper panel we plot the fraction of nodes in 
the largest connected component (LCC, denoted $q$)
as a function of the fraction of removed nodes, denoted $\rho$. We see that compared to HD and CI the CoreHD algorithm works the best by a large margin, breaking the network into small component with size smaller than $0.01 N$ after removing
fraction of only $0.1846$ of nodes. While CI and HD need to 
remove fraction $0.2014$, and $0.2225$ of nodes respectively. This is compared to the close-to-optimal performance of the iterative message passing BPD that needs to remove fraction $0.1780$ of nodes, and to the theoretical prediction for the asymptotically optimal value $0.1753$~\cite{zhou2013spin,altarelli2013optimizing,guggiola2015minimal,braunstein2016network,qin2016spin}.

We also see from the figure that the fraction of nodes in the LCC obtained by CoreHD 
encounters a first order transition when $\rho_{\mathrm{dec}}=0.1831$, this is because at this point (just at the beginning of the discontinuity) the 
remaining network becomes a forest.  The greedy tree-breaking procedure then quickly
breaks the forest into small components.
While the other algorithms do not have this phenomenon, the size of the LCC goes to zero continuously. In the lower panel of
Fig.~\ref{fig:ovl} we plot the fraction $q$ of nodes in the 
$2$-core as a function of $\rho$.
We can see that for CoreHD, $q$ reaches zero at $\rho=0.1831$ indicating that the remaining network
contains  no loop, thus is a forest. While for other algorithms the $2$-core remains extensive until the network is dismantled. On a larger ER random network with $N=10^6$, $c=3.5$, the difference between the sizes of decycling and the dismantling sets the CoreHD algorithm finds is not distinguishable within the precision of 4 valid digits and is $0.1830$ for both. Note that this result is (slightly) better than yet another approach suggested recently in the literature \cite{Clusella-etal-2016} that achieves $0.1838$ with an algorithm still considerably more involved than CoreHD. 

\begin{figure}[t]
  \begin{center}
    \includegraphics[angle=270,width=0.35\textwidth]{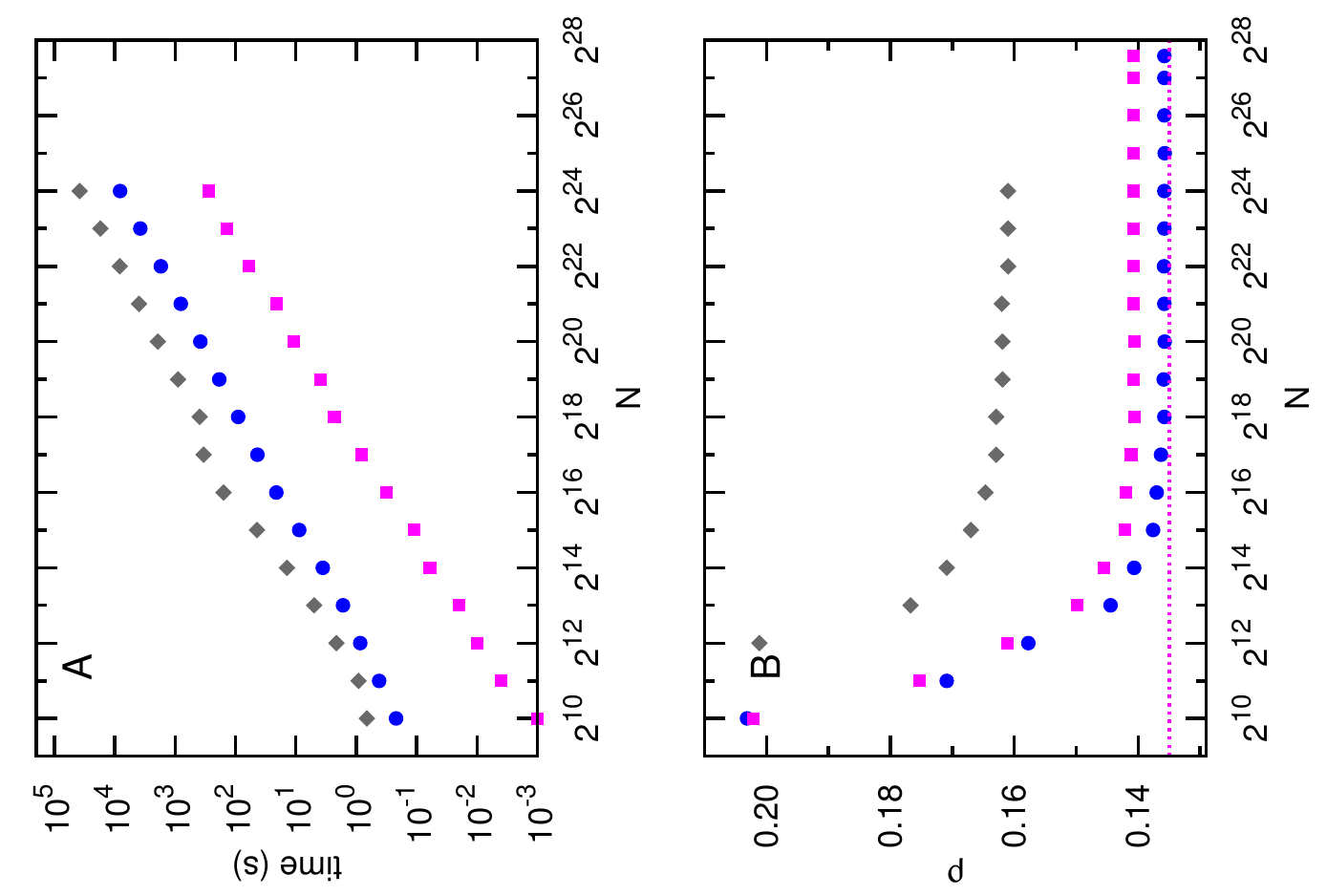}
  \end{center}
  \caption{\label{fig:timecomp}
   Performance of the CoreHD algorithm (magenta squares) and its comparison with the BPD algorithm (blue circles) and the CI algorithm ($\ell=4$, grey diamonds) on ER networks of average degree $c=3$ and size $N$. (A) The relationship between the total running time $\tau$ and
    $N$. The simulation results are obtained on a relatively
    old desktop computer (Intel-6300, $1.86$ GHz, $2$ GB memory). 
    (B) The relationship between the fraction $\rho$ of removed nodes and $N$. 
    The dotted horizontal line denotes the theoretically predicted minimum value. 
  }
\end{figure}

Besides performing much better than CI, the CoreHD is also much faster: the $2$-core of the
network can be computed efficiently using a leaf-removal process with 
$O(N)$ operations. After deleting a node, one only needs to update the $2$-core, which requires on average $O(1)$ operations
in sparse networks, and is clearly much faster than updating the CI score. Actually, in sparse networks when the size of the $2$-core is much smaller
than the size of the network, CoreHD is even faster than the HD algorithm which removes one by one nodes from the whole network.

\begin{figure*}[t]
 \begin{center}
    \includegraphics[angle=270,width=0.9\textwidth]{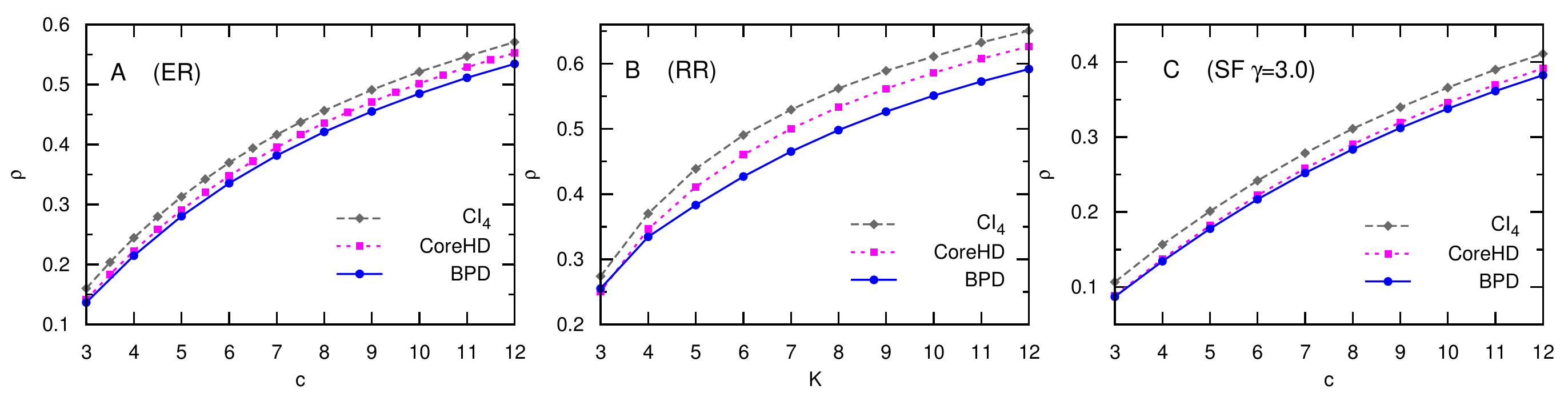}
  \end{center}
  \caption{\label{fig:ERaRRaSF}
    Fraction $\rho$ of removed nodes for (A) Erd\H{o}s-R\'enyi (ER) random networks of mean
    degree $c$, (B) Regular Random (RR) networks of degree $K$, and (C) Scale Free (SF) networks of average degree $c$ and decay exponent
    $\gamma=3.0$ generated as in~\cite{Goh-Kahng-Kim-2001}. Each data point obtained by
    CoreHD is the over $96$ instances
    of size $N=10^5$. The results of CI$_4$ and the results of BPD are
    from~\cite{mugisha2016identifying}. In BPD and CI$_4$, at each iteration a fraction $f$ of nodes are removed (with $f=0.01$ for BPD and $f=0.001$ for CI$_4$, decreasing $f$ does not improve the performance visibly), while in CoreHD nodes are removed one by one.
    }
 \end{figure*}

\begin{table*}[ht]
  \caption{
    \label{tab:realnet}
    Comparative results of the CoreHD method with CI and the BPD algorithm on a set of real-world
    network instances. $N$ and $M$ are the number of nodes and links of each
    network, respectively. The number of nodes deleted by CI, CoreHD,  and BPD are listed in the 4th, 5th, and 6th column.
    The CI and BPD results are from \cite{mugisha2016identifying}.
    The time (seconds) for dismantling is the running time of algorithms, i.e. with time for reading network from the data file excluded.
  }
  \begin{center}
%    \begin{footnotesize}
      \begin{tabular}{|l|r|r|rr|rrr|rrr|}
%      \begin{tabular}{|l|c|c|cc|ccc|ccc|}
        \hline 
          &  &  & \multicolumn{2}{c|}{decycling}& \multicolumn{3}{c|}{dismantling}&\multicolumn{3}{c|}{Time for dismantling} \\
        \hline
        Network  & $N$ & $M$ & CoreHD & \ \  \  \  BPD
        & CI &  CoreHD & \ \  \  \  BPD  & CI & CoreHD & BPD \\
        \hline 
        RoadEU \cite{Subelj-Bajec-2011}  &    $1177$ &     $1417$& ${\bf 90}$ & $91$   & $209$      & ${\bf 148 }$& $152$ &  $0.18$ & $<0.001$ & $0.1$ 
        \\
        PPI \cite{Bu-etal-2003}     &    $2361$ &     $6646$ & $365$ & $362$ 
        & $424$       & $357$  & ${350}$   & $0.91$ & $<0.001$ & $2.09$ 
        \\
        Grid \cite{Watts-Strogatz-1998}    &    $4941$ &     $6594$& $519$ & $512$ 
        & $476$     & $327$    & ${320}$   & $1.00$ & $<0.001$ &
        $0.66$ \\
        IntNet1 \cite{Leskovec-etal-2005}  &    $6474$ &    $12572$& ${217}$ & $215$ &
        $198$  & ${\bf {156}}$   & $161$  & $5.19$ & $<0.001$ & $11.32$ 
        \\ 
        Authors \cite{Leskovec-Kleinberg-Faloutsos-2007} &   $23133$ &    $93439$& ${\bf 8311}$ & $8317$ & 
        $3588$  & $2600$  &  ${2583}$ & $87.55$ & $0.09$ & $40.04$ 
         \\
        Citation \cite{Leskovec-etal-2005} &   $34546$ &   $420877$& $15489$ & $15390$ & 
        $14518$    & $13523$   & ${13454}$  & $4166$ & $0.2$ & $383.91$ 
        \\
        P2P \cite{Ripeanu-etal-2002}     &   $62586$ &   $147892$& $9557$ & $9285$ 
        & $10726$ & $9561$   & ${9292}$  & $520.59$  & $0.21$ & $50.24$ 
        \\
        Friend \cite{Cho-etal-2011}  &  $196591$ &   $950327$& $38911$ & $38831$ 
        & $32340$  & $27148$ & ${26696}$ & $5361$ & $1.37$ & $588.19$
        \\
        Email \cite{Leskovec-Kleinberg-Faloutsos-2007}   &  $265214$ &   $364481$& $1189$ & $1186$&
        $21465$  & $1070$     & ${1064}$ & $6678$ & $0.39$ & $151.57$
        \\
        WebPage \cite{Leskovec-etal-2009} &  $875713$ &  $4322051$& ${\bf 208509}$ & $ 208641$ & $106750$    & $51603$  & ${50878}$  &  $2275$ & $9.67$ & $2532$ 
        \\
        RoadTX \cite{Leskovec-etal-2009}  & $1379917$ &  $1921660$ & $243969$ & $239885$ &  $133763$ & $\textbf{20289}$ & $20676$ &  $273.69$ & $4.07$  & $421.15$  \\
        IntNet2 \cite{Leskovec-etal-2005} & $1696415$ & $11095298$& $229034$ & $228720$ & $144160$  & $73601$ & $73229$  & $19715$  & $35.84$ & $4243$
         \\
        \hline 
        %\hline
      \end{tabular} 
%    \end{footnotesize}
  \end{center}
\end{table*}  

The computational times for the CoreHG, CI and BPD algorithms as the system size grows are shown in Fig.~\ref{fig:timecomp} for ER network with mean degree $c=3$. The BPD algorithm performs slightly better than the CoreHD algorithm but it is much slower. For example, for
an ER network with $c=3$ and $N=2\times 10^8$, the solution obtained by CoreHD has relative
dismantling/decycling set size $\rho\approx 0.1407$ (computing time is $64$ minutes), which is only slightly
larger than the value of $\rho\approx 0.1357$ obtained by BPD (computing time is $23.5$ hours \cite{mugisha2016identifying}).
%To be compared to the optimum estimated to be at $\rho=0.1349$
%\cite{zhou2013spin,altarelli2013optimizing,guggiola2015minimal,braunstein2016network,qin2016spin}.
We note that in these experiments, in each step of removal, 
BPD and CI$_4$
remove $0.1\%$ of nodes (e.g., $10000$ nodes for $N=10^7$), while CoreHD
removes only $1$ node per step. Even this way the computational time of CoreHD is shorter than the time used for reading the network from the data file (edge-list format, using a c++ procedure).

Fig.~\ref{fig:ERaRRaSF} presents results for Erd\H{o}s-R\'enyi random graphs, regular random graphs, and scale-free random networks of varying average degree. In all cases CoreHD works better than CI and worse than BPD, with the best performance obtained for scale-free networks.
The good performance of CoreHD for the scale-free networks is of particular interest
because almost all real-world networks have a heavy-tailed degree-distribution. 

A set of experiments on real-world networks is presented in Tab.~\ref{tab:realnet}.
We list the fraction of nodes we need to remove in order to remove all cycles, and in order to break the network
into small components with size smaller than $0.01 N$. For dismantling, in addition to Algorithm~\ref{CoreHD} we do a refinement by inserting back some deleted nodes that do not increase the largest component size beyond the $0.01N$.  We can see that CoreHD works 
excellently for real-world network instances, giving decycling and dismantling sets very close to the state-of-art BPD and
much smaller than CI. It is also surprising to see that in some networks e.g. RoadEU, IntNet1 and RoadTX, CoreHD even outperforms BPD slightly.
Tab.~\ref{tab:realnet} clearly demonstrates the time superiority of CoreHD for real-world networks as compared with both CI and BPD. 
%For the larger networks it takes more time to load the dataset than to run the CoreHD algorithm.

\section{conclusion and discussions}

We have presented that iteratively removing nodes having the highest degree from the $2$-core of a network 
gives an ultra-fast while very efficient algorithm for decycling and dismantling of networks.
Our algorithm is so fast that its running time is shorter than the time of reading the network file.
%It means that there would be even not need to optimize the code for speed.

%Can one boost the performance of CI by restricting this algorithm on the $2$-core of the network? We checked that combining $2$-core reduction and CI into an algorithm called CoreCI does not improve the performance of the algorithm (see Appendix). It seems that simply deleting the nodes of highest degree from the $2$-core is the best local heuristic.

It is still surprising to us that such a simple algorithm could work much better than more sophisticated algorithms: 
We have tried running CI (see Appendix), adjacency matrix centrality on the $2$-core
of the network, and HD on $3$-core of the network, they are all slower but perform no better than CoreHD.
Our experiments also show that CoreHD outperforms centrality measures
using left and right eigenvector of the non-backtracking matrix~\cite{dogmat}, an idea that originally inspired us to propose the CoreHD algorithm. More detailed understanding of why this is the best performing strategy is let for future work. 
%We think the reason is that as there is no paramagnetic state for taking expansions, in the decycling and dismantling problems the relation between the non-backtracking matrix and the Belief Propagation are more complicated than in community detection problems. 
%In the future work we will be studying the detailed relation between the non-backtracking centrality and CoreHD.

On the real-world networks which typically have many short loops and motifs, decycling
is quite different from dismantling. A natural idea to generalize our CoreHD would be consider a factor graph treating short loops and motifs
as factors, then do CoreHD on the $2$-core of the factor graph. 
%We will put this into future work.

Finally, CoreHD can be generalized naturally to removal of the $k$-core, again running the adaptive DH heuristics on the $k$-core or the current graph. Comparison of this strategy to existing algorithms \cite{altarelli2013optimizing,pei2016collective} is in progress. 

\begin{acknowledgments}
H.J.Z was supported by the National Basic Research Program of China (grant number 2013CB932804), the National Natural Science Foundation of China (grant numbers 11121403 and 11225526), and the Knowledge Innovation Program of Chinese Academy of Sciences (No.~KJCX2-EW-J02).

\end{acknowledgments}

\bibliography{zp.bib}

\begin{thebibliography}{10}

\bibitem{zhou2013spin}
H.-J. Zhou.
\newblock Spin glass approach to the feedback vertex set problem.
\newblock {\em Eur. Phys. J. B}, 86:455, 2013.

\bibitem{altarelli2013optimizing}
F.~Altarelli, A.~Braunstein, L.~Dall’Asta, and R.~Zecchina.
\newblock Optimizing spread dynamics on graphs by message passing.
\newblock {\em Journal of Statistical Mechanics: Theory and Experiment},
  2013(09):P09011, 2013.

\bibitem{guggiola2015minimal}
A.~Guggiola and G.~Semerjian.
\newblock Minimal contagious sets in random regular graphs.
\newblock {\em Journal of Statistical Physics}, 158(2):300--358, 2015.

\bibitem{morone2015influence}
F.~Morone and H.~A. Makse.
\newblock Influence maximization in complex networks through optimal
  percolation.
\newblock {\em Nature}, 524:65--68, 2015.

\bibitem{mugisha2016identifying}
S.~Mugisha and H.-J. Zhou.
\newblock Identifying optimal targets of network attack by belief propagation.
\newblock {\em arXiv preprint arXiv:1603.05781}, 2016.

\bibitem{braunstein2016network}
A.~Braunstein, L.~Dall'Asta, G.~Semerjian, and L.~Zdeborov{\'a}.
\newblock Network dismantling.
\newblock {\em arXiv preprint arXiv:1603.08883}, 2016.

\bibitem{qin2016spin}
S.-M. Qin, Y.~Zeng, and H.-J. Zhou.
\newblock Spin glass phase transitions in the random feedback vertex set
  problem.
\newblock {\em arXiv preprint arXiv:1603.09032}, 2016.

\bibitem{Clusella-etal-2016}
P.~Clusella, P.~Grassberger, F.~J. {P\'erez-Reche}, and A.~Politi.
\newblock Immunization and targeted destruction of networks using explosive
  percolation.
\newblock arXiv:1604.00073, 2016.

\bibitem{karp1972reducibility}
R.~M. Karp.
\newblock Reducibility among combinatorial problems.
\newblock In {\em Complexity of computer computations}, pages 85--103.
  Springer, 1972.

\bibitem{janson2008dismantling}
S.~Janson and A.~Thomason.
\newblock Dismantling sparse random graphs.
\newblock {\em Combinatorics, Probability and Computing}, 17(02):259--264,
  2008.

\bibitem{Richardson-Domingos-2002}
M.~Richardson and P.~Domingos.
\newblock Mining knowledge-sharing sites for viral marketing.
\newblock In {\em Proceedings of 8th ACM SIGKDD International Conference on
  Knowledge Discovery and Data Mining}, pages 61--70, New York, NY, 2002. ACM.

\bibitem{Kempe-etal-2015}
D.~Kempe, J.~Kleinberg, and E.~Tardos.
\newblock Maximizing the spread of influence through a social network.
\newblock {\em Theory of Computing}, 11:105--147, 2015.

\bibitem{jung2012irie}
K.~Jung, W.~Heo, and W.~Chen.
\newblock Irie: Scalable and robust influence maximization in social networks.
\newblock In {\em 2012 IEEE 12th International Conference on Data Mining},
  pages 918--923. IEEE, 2012.

\bibitem{borgs2014maximizing}
C.~Borgs, M.~Brautbar, J.~Chayes, and B.~Lucier.
\newblock Maximizing social influence in nearly optimal time.
\newblock In {\em Proceedings of the Twenty-Fifth Annual ACM-SIAM Symposium on
  Discrete Algorithms}, pages 946--957. Society for Industrial and Applied
  Mathematics, 2014.

\bibitem{albert2000error}
R.~Albert, H.~Jeong, and A.-L. Barab{\'a}si.
\newblock Error and attack tolerance of complex networks.
\newblock {\em nature}, 406(6794):378--382, 2000.

\bibitem{cohen2001breakdown}
R.~Cohen, K.~Erez, D.~{ben-Avraham}, and S.~Havlin.
\newblock Breakdown of the internet under intentional attack.
\newblock {\em Physical review letters}, 86(16):3682, 2001.

\bibitem{bau2002decycling}
S.~Bau, N.~C. Wormald, and S.~Zhou.
\newblock Decycling numbers of random regular graphs.
\newblock {\em Random Structures \& Algorithms}, 21(3-4):397--413, 2002.

\bibitem{Goh-Kahng-Kim-2001}
K.-I. Goh, B.~Kahng, and D.~Kim.
\newblock Universal behavior of load distribution in scale-free networks.
\newblock {\em Phys. Rev. Lett.}, 87:278701, 2001.

\bibitem{Subelj-Bajec-2011}
L.~\v{S}ubelj and M.~Bajec.
\newblock Robust network community detection using balanced propagation.
\newblock {\em Eur. Phys. J. B}, 81:353--362, 2011.

\bibitem{Bu-etal-2003}
D.~Bu, Y.~Zhao, L.~Cai, H.~Xue, X.~Zhu, H.~Lu, J.~Zhang, S.~Sun, L.~Ling,
  N.~Zhang, G.~Li, and R.~Chen.
\newblock Topological structure analysis of the protein-protein interaction
  network in budding yeast.
\newblock {\em Nucleic Acids Res.}, 31:2443--2450, 2003.

\bibitem{Watts-Strogatz-1998}
D.~J. Watts and S.~H. Strogatz.
\newblock Collective dynamics of 'small-world' netowrks.
\newblock {\em Nature}, 393:440--442, 1998.

\bibitem{Leskovec-etal-2005}
J.~Leskovec, J.~Kleinberg, and C.~Faloutsos.
\newblock Graphs over time: densification laws, shrinking diameters and
  possible explanations.
\newblock In {\em Proceedings of the eleventh ACM SIGKDD international
  conference on Knowledge discovery in data mining}, pages 177--187. ACM, New
  York, 2005.

\bibitem{Leskovec-Kleinberg-Faloutsos-2007}
J.~Leskovec, J.~Kleinberg, and C.~Faloutsos.
\newblock Graph evolution: Densification and shrinking diameters.
\newblock {\em ACM Transactions on Knowledge Discovery from Data}, 1:2, 2007.

\bibitem{Ripeanu-etal-2002}
M.~Ripeanu, I.~Foster, and A.~Iamnitchi.
\newblock Mapping the gnutella network: Properties of large-scale peer-to-peer
  systems and implications for system design.
\newblock {\em IEEE Internet Comput.}, 6:50--57, 2002.

\bibitem{Cho-etal-2011}
E.~Cho, S.~A. Myers, and J.~Leskovec.
\newblock Friendship and mobility: User movement in localation-based social
  networks.
\newblock In {\em ACM SIGKDD International Conference o Knowledge Discovery and
  Data Mining}, pages 1082--1090, San Diego, CA, USA, 2011.

\bibitem{Leskovec-etal-2009}
J.~Leskovec, K.~J. Lang, A.~Dasgupta, and M.~W. Mahoney.
\newblock Community structure in large networks: Natural cluster sizes and the
  absence of large well-defined clusters.
\newblock {\em Internet Math.}, 6:29--123, 2009.

\bibitem{dogmat}
P.~Zhang.
\newblock Nonbacktracking operator for the ising model and its applications in
  systems with multiple states.
\newblock {\em Phys. Rev. E}, 91:042120, Apr 2015.

\bibitem{pei2016collective}
S.~Pei, X.~Teng, J.~Shaman, F.~Morone, and H.~A. Makse.
\newblock Collective influence maximization in threshold models of information
  cascading with first-order transitions.
\newblock {\em arXiv preprint arXiv:1606.02739}, 2016.

\bibitem{Krzakala2013}
F.~Krzakala, C.~Moore, E.~Mossel, J.~Neeman, A.~Sly, L.~Zdeborov\'a, and
  P.~Zhang.
\newblock Spectral redemption in clustering sparse networks.
\newblock {\em Proc. Natl. Acad. Sci. USA}, 110(52):20935--20940, 2013.

\end{thebibliography}

\begin{appendix}

\section{Greedy Tree Breaking and Refinement by Insertion}
\label{app:TBandRefine}

Optimally breaking a forest into small components can be solved 
in polynomial time \cite{janson2008dismantling}. Empirically a greedy tree-breaking procedure works very well. In such a greedy dynamics we iteratively 
find and remove the node which leads to the largest drop in 
the size of the largest connected component. 

In more details, the largest component caused by removal of each 
node in a tree can be computed iteratively (see, e.g.~\cite{mugisha2016identifying,braunstein2016network}). 
Starting from a leaf, each node sends a message to each of its 
neighbors, reporting the largest component caused by removing the 
edge between them. After the messages arrive at the root of the 
tree, we can then easily identify the node 
such that its removal decreases maximally the component 
size.

For the refinement, we also use a simple greedy strategy to 
insert back some of the removed nodes~\cite{mugisha2016identifying,braunstein2016network}.
In each step of re-insertion, we calculate the increase of
the component size after the insertion of a node, and then 
identify the node which gives the smallest increase.

\section{Dangling-tree problem of the CI index}
\label{app:CIproblem}

The collective influence index was proposed in \cite{morone2015influence} as a measure of node's importance in influence spreading. At a given level $\ell$ the CI index of a node $i$ is defined as
\begin{equation}
\label{eq:CI}
	\textrm{CI}_{\ell}(i)=(d_i-1) \sum_{j\in\partial^{\ell}_i}
    (d_j-1)\; ,
\end{equation}
where $d_i$ is the degree of node $i$ in the remaining network, and $\partial^{\ell}_i$ denotes the set of nodes that are
at distance $\ell$ from node $i$.
In the CI algorithm, a small fraction $f$ (e.g., $f=0.001$) of
nodes with the highest CI values are removed 
from the network and then the CI indices of the remaining nodes
are updated. The authors of \cite{morone2015influence} claimed that the CI$_{\ell}(i)$ approximates the 
eigenvector of the non-backtracking operator \cite{Krzakala2013}.

However we can see immediately that CI has a drawback which does not reflect the functioning of the non-bactracking operator. We illustrate this in an example network shown in Fig.~\ref{fig:dtree}. Without loss of generality let us consider
$\ell=2$, then it is easy to see that the node $i$ of this figure has  $\mathrm{CI}_2(i)>0$ and in some cases can be larger than the CI indices of the other nodes.
So the CI algorithm may say
node $i$ is more important to remove first, as its removal decreases mostly the eigenvalue of
the non-backtracking matrix. After a moment of thought we see that this conclusion
is not correct, as removing node $i$ does not change the eigenvalue of the
non-backtracking matrix at all, because the eigenvalue of the non-backtracking matrix 
is the same as the $2$-core of the network, while node $i$ does not belong to the 
$2$-core of the network.

\section{Comparing CoreHD and CoreCI}
\label{app:CoreHDvsCoreCI}

\begin{table*}[h]
\caption{
\label{tab:CoreHDaCI}
Comparing the dismantling performance of CoreHD and CoreCI on ER, RR, and SF random networks. Each data point is the mean and standard deviation of $\rho$ (the fraction of deleted nodes) over $96$ dismantling solutions obtained by CoreHD or CoreCI on $96$ independent network instances of size $N=10^5$ and mean degree $c$ (ER and SF) or degree $K$ (RR). The ball radius of CoreCI is fixed to $\ell=4$. The SF network instances are generated by the static method~\cite{Goh-Kahng-Kim-2001}.
}
\begin{center}
%    \begin{footnotesize}
      \begin{tabular}{|rll|rll|rll|}
        \hline \hline
        \multicolumn{3}{|c|}{ER} &
        \multicolumn{3}{c|}{RR} & \multicolumn{3}{c|}{SF ($\gamma=3.0$)}\\
        \hline
        $c$ & CoreHD & CoreCI & $K$ & CoreHD & CoreCI &
        $c$ & CoreHD & CoreCI \\
        \hline 
	$3.0$ & $0.1413(3)$ & $0.1427(3)$ & $3$ & $0.25043(3)$ & $0.2539(2)$ & $3.0$ & $0.0886(3)$ & $0.0893(3)$ \\
    $4.0$ & $0.2226(4)$ & $0.2249(4)$ & $4$ & $0.3464(2)$ & $0.3564(3)$ & $4.0$ & $0.1373(4)$ & $0.1383(4)$\\
    $5.0$ & $0.2908(4)$ & $0.2937(4)$ & $5$ & $0.4110(2)$ & $0.4239(3)$ & $5.0$ & $0.1820(5)$ & $0.1833(5)$\\
    $6.0$ & $0.3476(4)$ & $0.3509(4)$ & $6$ & $0.4605(3)$ & $0.4733(3)$ & $6.0$ & $0.2222(5)$ & $0.2237(5)$ \\
    $7.0$ & $0.3954(4)$ & $0.3990(4)$ & $7$ & $0.5004(3)$ & $0.5128(3)$ & $7.0$ & $0.2582(5)$ & $0.2560(5)$\\
    $8.0$ & $0.4361(4)$ & $0.4400(5)$ & $8$ & $0.5335(3)$ & $0.5455(3)$ & $8.0$ & $0.2906(6)$ & $0.2925(6)$\\
    $9.0$ & $0.4712(4)$ & $0.4752(5)$ & $9$ & $0.5617(3)$ & $0.5733(3)$ & $9.0$ & $0.3196(5)$ & $0.3217(5)$\\
    $10.0$ & $0.5018(4)$ & $0.5060(4)$ & $10$ & $0.5861(3)$ & $0.5974(4)$ & $10.0$ & $0.3460(6)$ & $0.3481(6)$ \\
    $11.0$ & $0.5288(4)$ & $0.5330(4)$ & $11$ & $0.6075(3)$ & $0.6182(4)$ & $11.0$ & $0.3699(6)$ & $0.3723(6)$\\
    $12.0$ & $0.5527(4)$ & $0.5571(4)$ & $12$ & $0.6264(3)$ & $0.6367(3)$ & $12.0$ & $0.3918(6)$ & $0.3943(6)$ \\
         \hline 
        \hline
      \end{tabular} 
%    \end{footnotesize}
  \end{center}
\end{table*}  
Since performing node deletion on the network $2$-core is the key of CoreHD's good performance, it is natural to expect that the CI algorithm can also be improved by adding the $2$-core reduction process. To confirm this, we implement an extended CI algorithm (named as CoreCI) as follows. At each elementary node removal step, (1) the $2$-core of the remaining network is obtained by cutting leaves recursively as in CoreHD, and then (2) the CI index of each node in the $2$-core is computed by considering only nodes and links within this $2$-core, and finally (3) a node with the highest CI index is deleted from the $2$-core. Similar to CoreHD and BPD, after a forest is produced by CoreCI, we then perform a greedy tree-breaking process if necessary and then re-insert some nodes back to the network as long as the size of the largest connected component is still below the threshold value of (say) $0.01 N$.

We indeed observe that CoreCI performs considerably better than the original CI algorithm. However it does not outperform CoreHD. We list in Table~\ref{tab:CoreHDaCI} the comparative results of CoreHD versus CoreCI on ER, RR, and SF random networks. Notice that the fractions $\rho$ of deleted nodes by CoreHD and CoreCI are very close to each other, with CoreHD performs slightly better. These results clearly demonstrate that the CI index is not a better indicator of node importance than the degree in the $2$-core. Because repeatedly computing the CI indices within the $2$-core is still very time-consuming, we recommend CoreHD rather than CoreCI as an efficient heuristic for practical applications.

\end{appendix}

\end{document}